



\input lanlmac
\input amssym
\input epsf

\newcount\figno
\figno=0
\def\fig#1#2#3{
\par\begingroup\parindent=0pt\leftskip=1cm\rightskip=1cm\parindent=0pt
\baselineskip=13pt
\global\advance\figno by 1
\midinsert
\epsfxsize=#3
\centerline{\epsfbox{#2}}
\vskip 12pt
{\bf Fig. \the\figno:~~} #1 \par
\endinsert\endgroup\par
}
\def\figlabel#1{\xdef#1{\the\figno}}
\newdimen\tableauside\tableauside=1.0ex
\newdimen\tableaurule\tableaurule=0.4pt
\newdimen\tableaustep
\def\phantomhrule#1{\hbox{\vbox to0pt{\hrule height\tableaurule
width#1\vss}}}
\def\phantomvrule#1{\vbox{\hbox to0pt{\vrule width\tableaurule
height#1\hss}}}
\def\sqr{\vbox{%
  \phantomhrule\tableaustep

\hbox{\phantomvrule\tableaustep\kern\tableaustep\phantomvrule\tableaustep}%
  \hbox{\vbox{\phantomhrule\tableauside}\kern-\tableaurule}}}
\def\squares#1{\hbox{\count0=#1\noindent\loop\sqr
  \advance\count0 by-1 \ifnum\count0>0\repeat}}
\def\tableau#1{\vcenter{\offinterlineskip
  \tableaustep=\tableauside\advance\tableaustep by-\tableaurule
  \kern\normallineskip\hbox
    {\kern\normallineskip\vbox
      {\gettableau#1 0 }%
     \kern\normallineskip\kern\tableaurule}%
  \kern\normallineskip\kern\tableaurule}}
\def\gettableau#1 {\ifnum#1=0\let\next=\null\else
  \squares{#1}\let\next=\gettableau\fi\next}

\tableauside=1.0ex
\tableaurule=0.4pt


\def\ep{\epsilon}

\def\hf{{1\over 2}}
\def\qu{{1\over 4}}

\def\o{\over}

\def\del{\partial}

\def\lf{\left}
\def\ri{\right}

\def\bt{\beta}

\def\Ga{\Gamma}

\def\Om{\Omega}

\def\rt#1{\sqrt{#1}}

\def\sitarel#1#2{\mathrel{\mathop{\kern0pt #1}\limits_{#2}}}

\lref\ColemanAW{
  S.~R.~Coleman and F.~De Luccia,
  ``Gravitational Effects On And Of Vacuum Decay,''
  Phys.\ Rev.\  D {\bf 21}, 3305 (1980).
}
\lref\SenQA{
  A.~Sen,
  ``Time and tachyon,''
  Int.\ J.\ Mod.\ Phys.\  A {\bf 18}, 4869 (2003)
  [arXiv:hep-th/0209122].
}
\lref\SenMV{
  A.~Sen,
  ``Remarks on tachyon driven cosmology,''
  Phys.\ Scripta {\bf T117}, 70 (2005)
  [arXiv:hep-th/0312153].
}
\lref\SasakuraTQ{
  N.~Sasakura,
  ``A de Sitter thick domain wall solution by elliptic functions,''
  JHEP {\bf 0202}, 026 (2002)
  [arXiv:hep-th/0201130].
}
\lref\HartleAI{
  J.~B.~Hartle and S.~W.~Hawking,
  Phys.\ Rev.\  D {\bf 28}, 2960 (1983).
}
\lref\FlanaganDY{
  E.~E.~Flanagan, S.~H.~H.~Tye and I.~Wasserman,
  ``Brane world models with bulk scalar fields,''
  Phys.\ Lett.\  B {\bf 522}, 155 (2001)
  [arXiv:hep-th/0110070].
}
\lref\SusskindKW{
  L.~Susskind,
  ``The anthropic landscape of string theory,''
  arXiv:hep-th/0302219.
}
\lref\ZamolodchikovXS{
  A.~Zamolodchikov and A.~Zamolodchikov,
  ``Decay of metastable vacuum in Liouville gravity,''
  arXiv:hep-th/0608196.
}
\lref\GutperleAI{
  M.~Gutperle and A.~Strominger,
  ``Spacelike branes,''
  JHEP {\bf 0204}, 018 (2002)
  [arXiv:hep-th/0202210].
}

\Title{             
}
{\vbox{
\centerline{Instanton Solution in Tachyon Cosmology}
}}

\vskip .2in

\centerline{Kazumi Okuyama}
\vskip5mm
\centerline{Department of Physics, Shinshu University}
\centerline{Matsumoto 390-8621, Japan}
\centerline{\tt kazumi@azusa.shinshu-u.ac.jp}
\vskip .2in

\vskip 3cm
\noindent

We find an exact classical solution 
in Euclidean gravity coupled to a scalar field
with a particular form of potential commonly used in tachyon cosmology.
This solution represents a tunneling between two vacua.   

\Date{May 2007}
\vfill
\vfill

\newsec{Introduction}
There is mounting evidence that string theory has a
landscape of vacua \SusskindKW.
In this picture,
it is likely that the universe experiences
a tunneling to a nearby vacuum.
As a first step, it is important to understand the
tunneling process in the low energy gravity description.
For instance, it is well known that
the false vacuum decay by bubble nucleation
is described by the Coleman-De Luccia instanton \ColemanAW.
However, the instanton solution is known only for a 
special circumstances where the thin-wall approximation is
valid\foot{See
\ZamolodchikovXS\ for a study of bubble nucleation in 2d quantum gravity.}.
It is desirable to have an analytic control
over the behavior of such instantons.

In this paper, we take a modest step toward this goal. We consider
a special form of scalar potential
\eqn\Vtach{
V(\phi)={1\o\cosh \phi}~,
}
and find a tunneling solution interpolating two vacua at $\phi=\pm\infty$.
This particular form of potential naturally appears in
the study of cosmology in the presence of unstable
D-brane \refs{\SenMV,\SenQA}. The scalar field $\phi$ is identified as
the open string tachyon $T$ on the D-brane.
In such a scenario, it is usually assumed that the kinetic
term of tachyon field is given by the Born-Infeld form.
However, this leads to a non-linear dynamics of tachyon field
and it is beyond our capability of analytic control.
Therefore, for simplicity we assume that the kinetic term of
scalar field is canonical.
\newsec{Instanton Solution}
In this section, we construct an exact solution of
the Euclidean gravity coupled to a scalar field
with a special choice of the scalar potential $V(\phi)$.
The Euclidean action of the system is given by
\eqn\SEuc{
S=\int d^4x\rt{g}\lf(-{R\o16\pi G_N}+\hf g^{\mu\nu}
\del_{\mu}\phi\del_{\nu}\phi+V(\phi)\ri)~.
}
We look for an $SO(4)$ symmetric solution with the ansatz
\eqn\matricphi{
ds^2=d\tau^2+a(\tau)^2d\Om^2_3,\quad
\phi=\phi(\tau)~,
} 
where $d\Om_3^2$ is the metric of round 3-sphere of unit radius.
The equations of motion are
\eqn\Vdeleq{
\ddot{\phi}+{3\dot{a}\o a}\dot{\phi}-{dV(\phi)\o d\phi}=0~,
}
and 
\eqn\Habbles{
{\dot{a}^2\o a^2}={1\o a^2}+{2\o3M_{pl}^2}\lf(\hf\dot{\phi}^2-V(\phi)\ri)~.
}
Here we defined the Planck mass $M_{pl}$ as
\eqn\Mplanck{
M_{pl}^2={1\o4\pi G_N}~.
}

To solve \Vdeleq\ and \Habbles, we follow the strategy
of \refs{\SasakuraTQ,\FlanaganDY}. Namely, we further assume
that $\dot{\phi}$ and $a$ are related by some function $f(a)$
\eqn\fadef{
\dot{\phi}^2=a{df(a)\o da}~.
}
Multiplying $\dot{\phi}$ to \Vdeleq\ and use \fadef,
we can easily see that \Vdeleq\ becomes total derivative with respect to $\tau$
\eqn\dtauone{
{d\o d\tau}\lf(\hf\dot{\phi}^2-V(\phi)+3f(a)\ri)=0~.
}
Integrating this equation and using \fadef, we find
\eqn\Vbyf{
V(\phi)={a\o2}{df(a)\o da}+3f(a)~.
}
Here we have included the integration constant
 in the definition of $f(a)$.
Plugging \Vbyf\ into \Habbles, we get a closed equation for $a$
\eqn\Adotf{
\dot{a}^2=1-{2\o M_{pl}^2} a^2f(a)~.
}

Now we have a freedom to choose $f(a)$. Each choice of $f(a)$
corresponds to a particular form of the scalar potential
obtained from \Vbyf\ and \fadef. As discussed in \SasakuraTQ,
the following choice of $f(a)$ leads to a solvable system 
\eqn\fachice{
f(a)={m^2M_{pl}^2\o2}\Big(1-\bt^2+\bt^2m^2a^2\Big)~.
}
Here $\bt$ is a dimensionless parameter and
$m$ is a mass parameter.

Now, let us first consider the solution for $a$.
When $f(a)$ is given by \fachice,  \Adotf\ becomes 
\eqn\adoteq{
\dot{a}^2=(1-m^2a^2)(1+\bt^2m^2a^2)~.
}
This equation is solved by
the Jacobi elliptic function ${\rm sn}(z,k)$\foot{
${\rm sn}(z,k)$ is the inverse of the elliptic integral
\eqn\sndef{
{\rm sn}^{-1}(z,k)=\int_0^z{dx\o\rt{(1-x^2)(1-k^2x^2)}}~.
}
} 
\eqn\solasn{
a(\tau)={1\o m}{\rm sn}(m\tau,i\bt)~.
}

Next consider the solution $\phi(\tau)$.
Plugging the form of $f(a)$ into \fadef, we find
\eqn\phidota{
\dot{\phi}^2=m^4M_{pl}^2\bt^2a^2~.
}
Again, this is solved by a combination of 
Jacobi elliptic functions
${\rm cn}(z,k)$ and ${\rm dn}(z,k)$
\eqn\solphi{
\phi(\tau)=M_{pl}\tanh^{-1}\lf({{\rm cn}(m\tau,i\bt)\o{\rm dn}(m\tau,i\bt)}\ri)
~.
}
Using the identity of Jacobi elliptic functions
\eqn\snid{
{\rm sn}^2(z,k)+{\rm cn}^2(z,k)=1,\quad
k^2{\rm sn}^2(z,k)+{\rm dn}^2(z,k)=1,
}
one can easily see that $a(\tau)$ and $\phi(\tau)$ are related by
\eqn\atophi{
(ma)^2={1\o2\bt^2}\lf(\bt^2-1+{1+\bt^2\o\cosh{2\phi\o M_{pl}}}\ri)~.
}
From \Vbyf, $V(\phi)$ is written in $a$ as
\eqn\Vphiina{
V(\phi)={m^2M_{pl}^2\o2}\Big[3(1-\bt^2)+4\bt^2(ma)^2\Big]~.
}
Finally, using \atophi\ and \Vphiina,
we obtain the form of scalar potential $V(\phi)$ 
for our choice of $f(a)$
\eqn\Vphiform{
V(\phi)=m^2M_{pl}^2\lf({1-\bt^2\o2}+{1+\bt^2\o \cosh{2\phi\o M_{pl}}}\ri)~.
}
As advertised, this is exactly the potential appeared in the
tachyon cosmology.

This potential has a maximum at $\phi=0$ with positive value
\eqn\Vmax{
V(\phi=0)=m^2M_{pl}^2{3+\bt^2\o2}>0 ~,
}
and two vacua at $\phi=\pm\infty$
\eqn\Vvac{
V(\phi=\pm\infty)=m^2M_{pl}^2{1-\bt^2\o2}~.
}
Note that the sign of vacuum energy at $\phi=\pm\infty$
depends on the value of $\bt$. 
For definiteness we consider the case $\bt=1$. Analysis of other cases
is straightforward.

\subsec{$\bt=1$ case}

When $\bt=1$, the scalar potential \Vphiform\ becomes
\eqn\Vbtone{
V(\phi)={2m^2M_{pl}^2\o \cosh{2\phi\o M_{pl}}}~.
}
$V(\phi)$ has an unstable tachyonic vacuum at $\phi=0$
and stable zero energy vacua  at $\phi=\pm\infty$.
Near $\phi=0$, $V(\phi)$ behaves as
\eqn\Vnearz{
V(\phi)\sim 2m^2M_{pl}^2-4m^2\phi^2,\qquad(|\phi|\ll M_{pl})~.
}
From this we see that the parameter $m$ sets the 
mass scale of the tachyon.

The solution for $a(\tau)$ and $\phi(\tau)$ for the $\bt=1$ case is
\eqn\solbtone{
a(\tau)={1\o m}{\rm sn}(m\tau,i),\quad
\phi(\tau)=M_{pl}\tanh^{-1}\lf({{\rm cn}(m\tau,i)\o {\rm dn}(m\tau,i)}\ri)~.
}
From the periodicity of Jacobi elliptic functions,
we can easily see that
\eqn\azero{
a(0)=0,\quad a(\tau_c/2)={1\o m},\quad a(\tau_c)=0~,
}
\eqn\phizero{
\phi(0)=+\infty,\quad \phi(\tau_c/2)=0,\quad \phi(\tau_c)=-\infty~,
}
where $\tau_c$ is given by
\eqn\taucdef{
\tau_c={2\o m}\int_0^1{dx\o\rt{1-x^4}}={1\o\rt{2\pi}m}\Ga\lf(\qu\ri)^2~.
}
Namely, this solution interpolates between two vacua
at $\phi=\pm\infty$, and at each vacuum the radius of $S^3$ shrinks to zero.
Since $a(\tau)$ behaves as
$a\sim\tau$ near $\tau=0$,
the metric \matricphi\ is smooth at $\tau=0$.
Also, the geometry is smooth
at the other end $\tau=\tau_c$, since $a(\tau)=a(\tau_c-\tau)$. 
Therefore, our solution is an $S^3$ bundle over the interval
$[0,\tau_c]$ with vanishing $S^3$'s at the boundaries.
It is clear that the topology of the total space is $S^4$ (see fig. 1).
The maximum radius of $S^3$ is achieved at the tachyonic vacuum
$\phi=0$, and its value is given by the inverse of tachyon mass:
$a(\tau_c/2)=1/m$.  
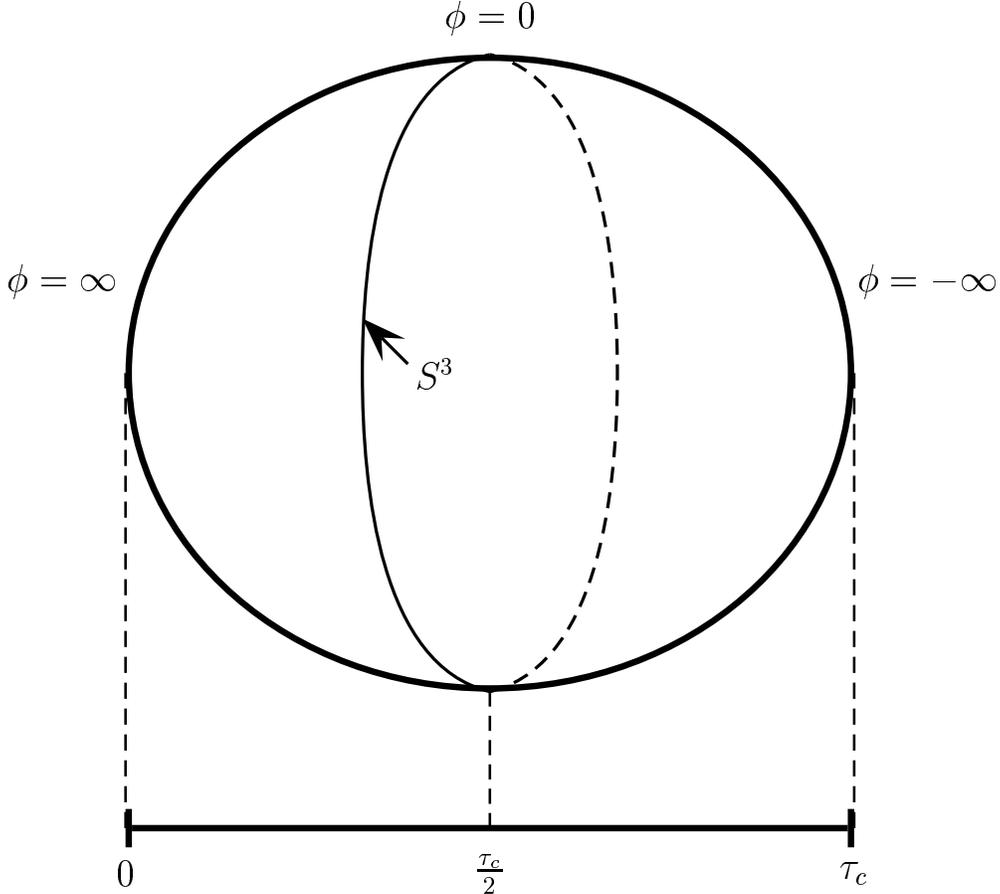
\fig{The instanton solution interpolating two vacua at $\phi=\pm\infty$.
$S^3$ is fibered over the interval $\tau\in[0,\tau_c]$.
The topology of this solution is $S^4$.}{3sphere.epsi}{11cm}

\newsec{Discussion}
We found an exact instanton solution of tachyon cosmology.
We emphasize that we did not use any approximation, such
as a thin-wall approximation.
Our solution can be used as a zero-th order
term for a more interesting case, {\it i.e.}
a bubble nucleation
in the false vacuum
\eqn\Vacdif{
V(\infty)-V(-\infty)=\ep\not=0~,
}
assuming that $\ep\ll V(0)$.
Also, it is interesting to
consider the half of our Euclidean solution
as the Hartle-Hawking 
no boundary state \HartleAI,
and glue the Lorentzian inflationary spacetime
at $\tau=\tau_c/2$.
It would be also interesting to consider the relation to
the S-brane solution \GutperleAI.
\vskip5mm
\noindent
{\bf Acknowledgment:} This work is supported in part by
JSPS Grant-in-Aid for Scientific Research \#19740135.

\listrefs
\bye